\documentclass[aps,prb,twocolumn,superscriptaddress]{revtex4}
\usepackage{epsf,graphicx}
\usepackage{amssymb}
\usepackage{amsmath}

\usepackage{color}
\usepackage{ulem}
\usepackage{booktabs}
\usepackage{multirow}
\begin{document}

\title{Pair density wave, unconventional superconductivity, and non-Fermi liquid quantum critical phase in frustrated Kondo lattice}
\author{Jialin Chen}
\affiliation{Beijing National Laboratory for Condensed Matter Physics and Institute of
Physics, Chinese Academy of Sciences, Beijing 100190, China}
\affiliation{University of Chinese Academy of Sciences, Beijing 100049, China}
\author{Jiangfan Wang}
\affiliation{School of Physics, Hangzhou Normal University,  Hangzhou, Zhejiang 311121, China}
\author{Yi-feng Yang}
\email[]{yifeng@iphy.ac.cn}
\affiliation{Beijing National Laboratory for Condensed Matter Physics and Institute of
Physics, Chinese Academy of Sciences, Beijing 100190, China}
\affiliation{University of Chinese Academy of Sciences, Beijing 100049, China}
\affiliation{Songshan Lake Materials Laboratory, Dongguan, Guangdong 523808, China}
\date{\today}

\begin{abstract}
Motivated by the recent discovery of an intermediate quantum critical phase between the antiferromagnetic order and the Fermi liquid in the frustrated Kondo lattice CePdAl, we study here a Kondo-Heisenberg chain with frustrated $J_1$-$J_2$ XXZ interactions among local spins using the density matrix renormalization group method. Our simulations reveal a global phase diagram with rich ground states including the antiferromagnetic order, the valence-bond-solid and bond-order-wave orders, the pair density wave state, the uniform superconducting state, and the Luttinger liquid state. We show that both the pair density wave and uniform superconductivity belong to the family of Luther-Emery liquids and may arise from pair instability of an intermediate quantum critical phase with medium Fermi volume in the presence of strong quantum fluctuations, while the Luttinger liquid has a large Fermi volume. Our work provides a comprehensive picture of the frustrated Kondo lattice physics at one dimension, and suggests a deep connection between the pair density wave, the unconventional superconductivity, and the non-Fermi liquid quantum critical phase.
\end{abstract}

\maketitle

\section{Introduction}
The competition between Kondo screening and the Rudermann-Kittel-Kasuya-Yosida (RKKY) interaction underlies the rich physics of Kondo lattice systems \cite{Yangn2017Quantum,Paschen2021Quantum}. Near the quantum critical point (QCP) \cite{Sachdev1999Quantum} from an antiferromagnetic (AFM) ground state to a Fermi liquid, anomalous non-Fermi-liquid (NFL) properties and unconventional superconductivity (SC) often emerge, as have been observed in YbRh$_2$Si$_2$ \cite{Custers2003Quantum, Paschen2004Hall}, CeRhIn$_5$ \cite{Park2006Hidden}, CeCu$_2$Si$_2$ \cite{Stockert2011Magnetically}, and CeCu$_{6-x}$Au$_x$ \cite{Schroder2000Onset}. Without frustration, the AFM QCP is believed to coincide with a sharp transition from small to large electron Fermi surfaces (FS) \cite{Rabello2001Locally, Paschen2004Hall, Friedemann2010Fermi, Shishido2005A, Knebel2008The}, while in frustrated Kondo lattices such as CePdAl \cite{Zhao2019Quantum,Majumder2022PRB}, the QCP was reported to expand into a finite region of NFL ground state. This quantum critical phase may arise from strong quantum fluctuations due to geometric or interaction frustrations or low dimensionality \cite{ Paschen2021Quantum, Coleman2010Frustration}, but its nature is not yet clarified. Previous theories proposed a weak-coupling state consisting of almost decoupled spinons and electrons with a small electron FS \cite{Pixley2014Quantum}, while recent work found a strong-coupling state with well-defined holon excitations as composite objects of spinons and holes and an electron FS of medium size \cite{Wang2021Nonlocal, Wang2022Z2}. In the latter theory, it has been proposed that the intermediate quantum critical phase could become unstable towards other exotic orders such as holon superconductivity, pair density wave (PDW), or hybridization wave. Investigations using more accurate methods are needed to settle this debate.

However, numerical calculations of two-dimensional (2D) frustrated Kondo lattices are difficult. Quantum Monte Carlo simulations often suffer from severe negative sign problems \cite{Sato2018Quantum, Motome2010Partial}. To proceed, we take the 1D Kondo lattice as a starting point, which may be solved using the well-developed density matrix renormalization group (DMRG) approach \cite{White1992Density, Stoudenmire2012Studying,Moukouri1995Density} and help to illuminate the basic physics of the 2D systems under doping \cite{Doniach1977The}. DMRG has been extensively applied to 1D Kondo lattice systems and revealed many important features such as an intermediate ferromagnetic phase for doping less than 0.5 \cite{McCulloch2002Interplay}, charge density wave (CDW) in doped Kondo chain \cite{Huang2019CDW}, the dimerization \cite{Xavier2003Dimerization,Xavier2008One,Huang2020Coupled} and bond-order-wave (BOW) order \cite{Huang2020Coupled} at quarter filling, and possible PDW state \cite{Berg2010Pair, Jaefari2012Fradkin, Dobry2013Inhomogeneous, Mann2020Topology, Zhang2022Pair}. It has also confirmed the existence of a small electron FS with the Fermi wave vector $k^\mathrm{S}_\mathrm{F} =\frac{n}{2}\pi$ ($n$ is the density of conduction electrons) at weak coupling and a large FS with $k^\mathrm{L}_\mathrm{F} =\frac{1+n}{2}\pi$ at strong Kondo coupling \cite{Moukouri1995Density, Shibata1996Friedel, Shibata1997Friedel, Tsunetsugu1997The, Shibata1999The, McCulloch2002Interplay, Xie2015Interplay, Khait2018Doped,Eidelstein2011Interplay}, but whether or not there exits an intermediate quantum critical phase and what its instability might lead to have not been well explored.
 
In this work, we study this issue based on the 1D Kondo-Heisenberg model with frustrated $J_1$-$J_2$ XXZ interactions using the infinite DMRG method. The nearest ($J_1$) and next-to-nearest ($J_2$) Heisenberg interactions are included to account for the magnetic frustration characterized by the parameter $Q=J_2/J_1$. A global phase diagram is constructed and reveals unambiguous evidences for intermediate strong-coupling phases over a wide parameter region. The ground states are found to be governed by pair correlations, causing a PDW state near the AFM phase boundary and a uniform SC state at larger Kondo coupling $J_\mathrm{K}$. We show that both the PDW and SC phases belong to Luther-Emery liquids and are eventually suppressed to enter a Luttinger liquid (LL) with a large Fermi volume as $J_\mathrm{K}$ increases. Our work reveals a deep connection between the pair density wave, the unconventional superconductivity, and the non-Fermi liquid quantum critical phase.

The paper is organized as follows. Section \ref{SecModel} introduces the model and the DMRG method used in our calculations. Section \ref{SecPhaseDiagram} gives a typical $Q$-$J_\mathrm{K}$ phase diagram for $n=0.8$ and $J_1=0.6$, showing all competing ground states including AFM, valence bond solid (VBS), BOW, PDW, SC, LL, and charge density wave (CDW) orders. In section \ref{SecDiss}, we clarify the Fermi volume evolution over the phase diagram and attribute the PDW and SC states to the pair instability of an intermediate quantum critical phase with a partially enlarged Fermi volume. The effects of electron density variation on the phase diagram and the Fermi volume are also discussed.

\begin{figure}[t]
	\begin{center}
		\includegraphics[width=8.0cm]{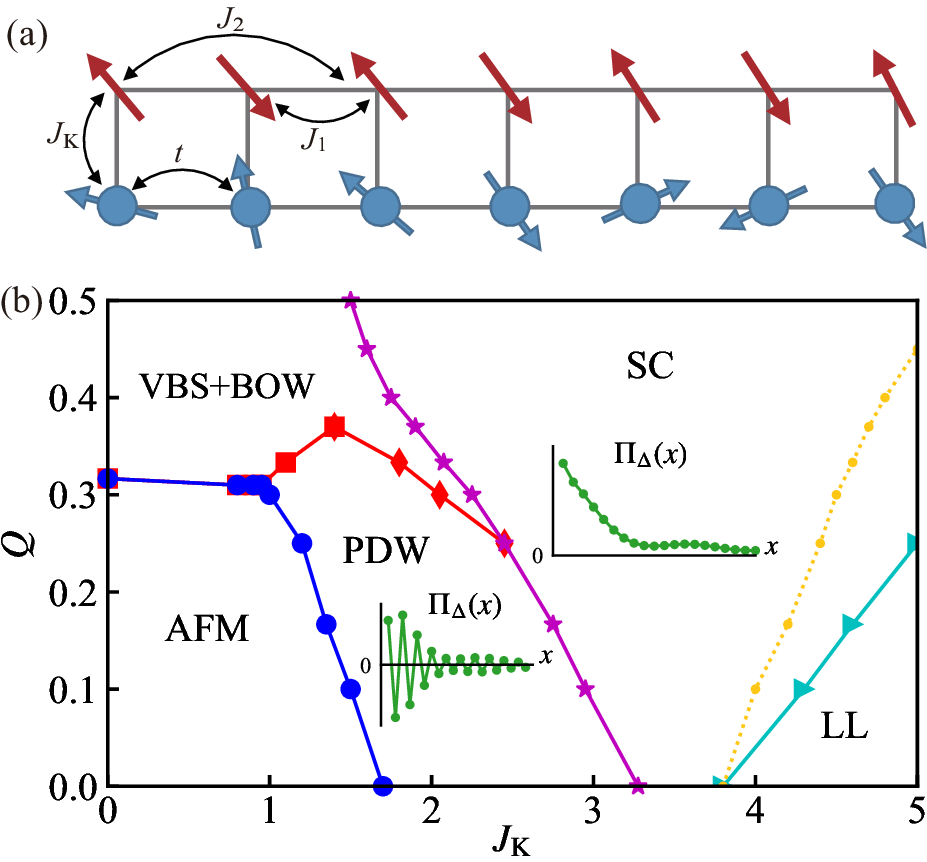}
	\end{center}
	\caption{(a) Illustration of the 1D Kondo-Heisenberg model with frustrated nearest-neighbor and next-nearest-neighbor spin interactions. The local spins (red arrows) and the conduction electrons (blue balls with arrows) are coupled through the Kondo coupling $J_\mathrm{K}$. (b) Typical phase diagram on $Q$-$J_\mathrm{K}$ plane based on calculations for $J_1=0.6$ and $n=0.8$. The insets illustrate the behaviors of the pair correlation $\Pi_\Delta(x)$ in PDW and SC states. The yellow dotted line separates the regions with medium (left) and large (right) Fermi volume, which is small in AFM, PDW, and VBS+BOW regions.}
	\label{fig1}
\end{figure}

\section{Model}{\label{SecModel}}
We start with the Kondo-Heisenberg Hamiltonian,
\begin{equation}{\label{Eq-Model}}
	\begin{split}
		H = &-t \sum_{\langle ij\rangle,\sigma} \left( c^\dagger_{i,\sigma} c_{j,\sigma} + \text{H.c.} \right) + J_\mathrm{K}\sum_i \mathbf{s}_i \cdot \mathbf{S}_i \\
		&+ J_1 \sum_{\langle ij\rangle} \left( S^x_i S^x_j + S^y_i S^y_j + \Delta_z S^z_i S^z_j \right) \\
		&+ J_2 \sum_{\langle\langle ij\rangle\rangle} \left( S^x_i S^x_j + S^y_i S^y_j + \Delta_z S^z_i S^z_j \right),
	\end{split}
\end{equation}
where $c^\dagger_{i,\sigma} $ creates an electron of spin $\sigma$ at site $i$, $t$ is the nearest-neighbor hopping amplitude, and $J_\mathrm{K}$ is the Kondo coupling between the local spin $\mathbf{S}_i$ and the conduction electron spin $\mathbf{s}_i=\sum_{\alpha\beta}c_{i,\alpha}^{\dag}\frac{\vec{\sigma}_{\alpha\beta}} {2}c_{i,\beta}$, where $\vec{\sigma}$ are the Pauli matrices. The Heisenberg exchange interactions $J_1$ and $J_2$ are both positive and describe AFM couplings between nearest-neighbor and next-nearest-neighbor local spins, respectively. Their ratio $Q=J_2/J_1$ reflects the strength of magnetic frustration. $\Delta_z$ gives the anisotropy of the exchange interactions along $z$-axis in the spin space and is set to 2 throughout this work. Hereafter, $t$ is set to unity. 

An intuitive illustration of the above frustrated Kondo-Heisenberg model (FKHM) is shown in Fig.~\ref{fig1}(a). We study the model using the infinite DMRG (iDMRG) method \cite{White1992Density, White1993Density, McCulloch2008Infinite, itensor, tenpy}, which is a variation method based on infinite matrix product states consisting of unit cells arranged along one dimension periodically\cite{McCulloch2008Infinite, Orus2014, Schollwock2011}. It is suitable for accurately determining the Luttinger parameters without being affected by the boundary effect. To account for possible periodicity of the ordered states (AFM, VBS, CDW, PDW) and spin correlations, we use different unit-cell sizes $L_\mathrm{u}=20$, 10, 10, 8 for the electron density $n=0.9$, 0.8, 0.6, 0.5, respectively, and ensure the consistency of the unit-cell sizes by checking with the finite-size DMRG. Since the total electron number $\sum_{i,\sigma} c^\dagger_{i,\sigma} c_{i,\sigma}$ and the $z$-component of the total spins $\sum_i (s^z_i + S^z_i)$ are both conserved, we use the U$(1)$ $\times$ U$(1)$ symmetry to accelerate the computation. The virtual bond dimension is chosen to be $\chi=4000$ in most cases, with which we can achieve a typical truncation error smaller than $4\times 10^{-6}$. For more challenging situations, a larger $\chi=8000$ is used. To ensure the convergence, we perform typically $80$ to $160$ sweeps for each calculation. More details of our simulations can be found in the appendix \ref{Appendix_A}.  

\section{The ground state phase diagram}{\label{SecPhaseDiagram}}
A typical phase diagram is given in Fig.~\ref{fig1}(b) based on our calculations for $J_1=0.6$ and $n=0.8$. The corresponding order parameters or characteristics are summarized in Table~\ref{tab_1}. For small $J_\mathrm{K}$, we obtain an AFM phase at small $Q$ and a VBS phase coexisting with a BOW order at large $Q$. There is a deconfined quantum critical point (DQCP) between the AFM and VBS/BOW phases that persists up to $J_\mathrm{K}\approx 1.0$ and then splits into an intermediate quantum critical region showing PDW order with a Fermi wave vector consistent with that of noninteracting conduction electrons. Further increasing $J_\mathrm{K}$ suppresses the PDW order and induces a first-order transition to a uniform SC phase accompanied with a jump in the Fermi wave vector. In the SC state, the Fermi volume is not sharply defined. It looks like to have a medium size and evolves gradually to the large Fermi volume for sufficiently large $J_\mathrm{K}$ (the yellow dotted line). Both the PDW and SC phases show characteristic properties of the Luther-Emery liquid. When the SC is destroyed at even larger $J_\mathrm{K}$, the system enters a Luttinger liquid (LL) phase \cite{Haldane1981Luttinger, Giamarchi2003Quantum}. The continuous evolution of the Fermi volume implies an intermediate phase in contrast to the weak-coupling Kondo breakdown scenario, which predicts a sharp transition from small to large Fermi volumes. The PDW and SC states cover the whole intermediate region and arise from the electronic instability due to AFM fluctuations. The pair correlations also penetrate into the AFM phase and the region of large Fermi volume.

\begin{table}[t] 
	\centering 	 
	\caption{Characteristics of all phases in the $Q$-$J_\mathrm{K}$ phase diagram of the frustrated Kondo-Heisenberg model. The last column gives the definition of the order parameters in the main text.} 
	\label{tab_1}
	\vspace{7pt} 
	\begin{tabular}{l|l|c } 
	    \hline\hline
		Phase & \qquad \qquad\qquad Characteristic & Definition \\ 
		\hline
		AFM & \qquad\qquad\qquad\qquad $O_\mathrm{AFM}$ & Eq.~(\ref{eq_AFM}) \\
		\hline
		VBS & \qquad\qquad\qquad\qquad $O_\mathrm{VBS}$ & Eq.~(\ref{eq_VBS}) \\
		+BOW & \qquad\qquad\qquad\qquad $O_\mathrm{BOW}$ & Eq.~(\ref{eq_BOW}) \\		
		\hline 
		\multirow{4}{*}{LEL} &~1. $|\Pi_\Delta(x)|\propto|x|^{-K_\Delta}$ and $K_\Delta<2$~   &Eq.~(\ref{eq_SCx}) \\		
							 &~2. $\Pi_n(x)\propto|x|^{-K_n}$    &  Eq.~(\ref{eq_nx})  \\
							 &~3. $K_\Delta * K_n \simeq  1$   &   \\
							 &~4. $|\Pi_S(x)|\propto e^{-\Delta_S x}$  and $\Delta_S>0$ & Eq.~(\ref{eq_Sx}) \\
		\hline 
		\multirow{1}{*}{PDW} &~Belongs to LEL and $\Pi_\Delta(q=\pi)$ dominates &   \\		                    
		\hline 
		\multirow{1}{*}{SC} &~Belongs to LEL and $\Pi_\Delta(q=0)$ dominates &\\
		\hline 
		\multirow{2}{*}{LL} &~1. Quasi-long ranged $\Pi_S(x)$ or $\Delta_S\simeq 0$ & Eq.~(\ref{eq_Sx}) \\ 
		                    &~2. Short ranged $\Pi_\Delta(x)$ or $K_\Delta > 2.0$ & Eq.~(\ref{eq_SCx}) \\
		\hline\hline 
	\end{tabular}
\end{table}

\subsection{AFM and VBS/BOW}
\begin{figure}[t]
	\begin{center}
		\includegraphics[width=8.0cm]{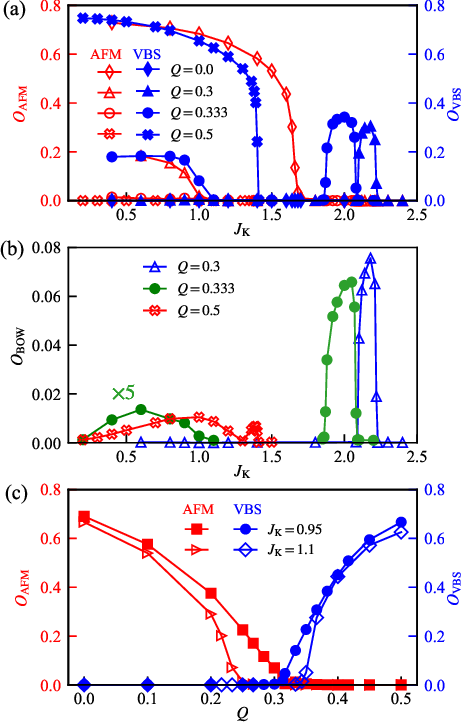}
	\end{center}
	\caption{Evolution of (a) $O_\mathrm{AFM}$ and $O_\mathrm{VBS}$, and (b) $O_\mathrm{BOW}$ with the Kondo coupling $J_\mathrm{K}$. (c) Evolution of $O_\mathrm{AFM}$ and $O_\mathrm{VBS}$ with the frustration parameter $Q$. In (b), $O_\mathrm{BOW}$ at $Q=0.333$ is magnified $5$ times for $J_\mathrm{K}<1.2$. Other parameters are $J_1=0.6$ and $n=0.8$.}
	\label{fig2}
\end{figure}

We discuss first the AFM and VBS/BOW phases. For $J_\mathrm{K}=0$, the conduction electrons and local spins are decoupled. The pure $J_1$-$J_2$ spin model has been extensively studied \cite{Haldane1982Critical, Nomura1994Critical} and, for $\Delta_z>1$, has an AFM ground state at small $Q$. Increasing $Q$ beyond a critical value $Q_c$ drives the system into a VBS phase. Numerically, the two phases can be identified by the following order parameters:
\begin{align}  
		O_{\mathrm{AFM}} &= \frac{2}{L} \sum_{l} (-1)^l \langle S^z_l\rangle, \label{eq_AFM}\\
		O_\mathrm{VBS} &= \frac{2}{L} \sum_{l} (-1)^l \langle \mathbf{S}_{l}\cdot \mathbf{S}_{l+1} \rangle.  \label{eq_VBS}
\end{align}
Their results are plotted in Fig.~\ref{fig2}(a) for several values of $Q$ as functions of $J_\mathrm{K}$. At $Q=0$, the AFM order is most stable. Its order parameter $O_\mathrm{AFM}$ is reduced with increasing Kondo coupling and vanishes continuously at the QCP $J_\mathrm{K}=1.7$. Introducing frustration suppresses the AFM order and drives the QCP to a smaller $J_\mathrm{K}=1.0$ at $Q=0.3$. The AFM order persistes up to $Q_c\simeq 0.32$ for $J_\mathrm{K}=0$, beyond which it is taken over by the VBS.

The VBS phase contains some peculiar feature that requires careful examination. As shown in Fig.~\ref{fig2}(a), it first appears at large $J_\mathrm{K}$ after the AFM is completely suppressed for $Q=0.3$. For $Q=0.333$, two separate regions with finite VBS order parameters are observed. For clarity, we refer them as the left and right VBS, respectively. The left one exists for $J_\mathrm{K} <1.1$ and corresponds to the VBS state of the pure spin model in the decoupling limit. It arises from the $J_1$-$J_2$ interaction among local spins and its order parameter decreases continuously as the Kondo coupling increases to $J_\mathrm{K}\approx1.1$. By contrast, the right VBS is only observed for a small range of $1.85<J_\mathrm{K}<2.1$ and vanishes much faster than the left one with increasing $J_\mathrm{K}$, which must have a distinct origin. We will see that it is actually inherited from $n=0.5$. For larger $Q=0.5$, the two regions merge together into a single region as shown in Fig.~\ref{fig1}(b).
  
The VBS order is always accompanied by a BOW order of itinerant electrons for $J_\mathrm{K}>0$, which characterizes the ordering of electron pairs on neighboring sites and may be quantified by \cite{Huang2020Coupled, Sengupta2002Bond, Farre2022Revealing}
\begin{equation}{\label{eq_BOW}}
O_\mathrm{BOW} = \frac{1}{L} \sum_{l\sigma} (-1)^l \langle c^\dagger_{l,\sigma} c_{l+1,\sigma}\rangle.
\end{equation}
The numerical results are plotted in Fig.~\ref{fig2}(b). We immediately draw two conclusions. First, direct comparison with Fig.~\ref{fig2}(a) indicates that the BOW order always coexists with the VBS order. Second, for fixed $Q$, the order parameter $O_\mathrm{BOW}$ of the right VBS phase can be much larger than that of the left one. These two observations together suggest that the BOW order arises from the VBS order of local spins through the Kondo coupling. The larger Kondo coupling in the right VBS/BOW region may be responsible for the much larger magnitude of the BOW order parameter, with the VBS order parameters being of similar magnitude in both regions.
 
The AFM and VBS phases break different symmetries and their quantum phase transition (QPT) at $J_\mathrm{K}=0$ has recently been shown to be beyond the conventional Landau-Ginzburg theory \cite{Mudry2019Quantum, Jiang2019Quantum, Roberts2019Quantum, Huang2019Emergent}. For finite $J_\mathrm{K}$, we may study both the effect of frustration $Q$ on the deconfined QCP and its stability when coupled to a Fermi sea \cite{Volkova2020Magnon}. This is shown in Fig.~\ref{fig2}(c), where we plot the AFM and VBS order parameters as functions of the frustration parameter $Q$ for two different values of $J_\mathrm{K}$. Within the limit of our numerical accuracy, the QCP remains stable even for $J_\mathrm{K} \simeq 0.95$, with the critical $Q_c \simeq 0.32$ only slightly reduced than that at $J_\mathrm{K}=0$. For slightly larger $J_\mathrm{K}=1.1$, the AFM and VBS phases are separated by a clear intermediate region, which is not present in the original spin model. 

\subsection{PDW and SC}
We demonstrate in this section that the intermediate region between AFM and VBS for $J_\mathrm{K} > 0.95$ exhibits an unusual PDW order or a uniform SC order. This is seen by introducing the pair correlation function:
\begin{equation}{\label{eq_SCx}}
\Pi_\Delta (x) = \left\langle\Delta^\dagger_x \Delta_0\right\rangle,
\end{equation}
where $\Delta^\dagger_l = ( c^\dagger_{l\uparrow} c^\dagger_{l+1\downarrow} - c^\dagger_{l\downarrow} c^\dagger_{l+1\uparrow} )$ represents the singlet-pair operator. In one dimension, it should exhibit quasi-long-range correlations characterized by a slow power-law decay, namely, $|\Pi_\Delta (x)| \propto |x|^{-K_\Delta} $. 

\begin{figure}[t]
	\begin{center}
		\includegraphics[width=8.6cm]{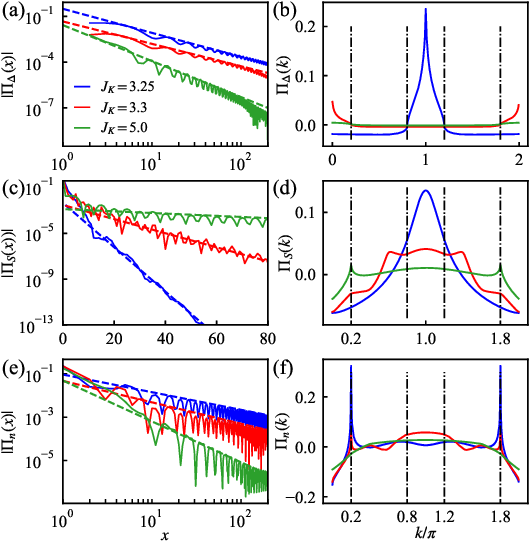}
	\end{center}
	\caption{Real space (left) and momentum space (right) variation of (a)(b) the singlet-pair correlation $\Pi_\Delta$,  (c)(d) the spin correlation $\Pi_S$, and (e)(f) the charge density correlation $\Pi_n$ for $J_\mathrm{K} = 3.25$, 3.3, and 5.0 at $Q=0$. The dashed lines represent the power-law fitting. Other parameters are $J_1=0.6$ and $n=0.8$.}
	\label{fig3}
\end{figure} 

Some typical results of the pair correlation function are plotted in Fig.~\ref{fig3}(a). We see a slow power-law decay in the amplitude $|\Pi_\Delta (x)|$ for both $J_\mathrm{K}=3.25$ (blue) and $3.3$ (red), indicating the presence of quasi-long-range singlet-pair correlations. However, the two results are actually different, which is seen after Fourier transformation to the momentum space. As shown in Fig.~\ref{fig3}(b), $\Pi_\Delta (k)$ is dominated by the peak at the momentum $k=\pi$ for $J_\mathrm{K}=3.25$ and at $k=0 ~ (\mathrm{mod }~ 2\pi)$ for $J_\mathrm{K}=3.3$. In real space, the sign of $\Pi_\Delta(x)$ oscillates for $J_\mathrm{K}=3.25$, while it remains positive for $J_\mathrm{K}=3.3$. The two situations therefore correspond to a PDW state with the wave length $\lambda_\mathrm{PDW}=2$ and a uniform SC state, as illustrated in the insets of Fig.~\ref{fig1}(b).
 
\begin{figure}[t]
	\begin{center}
		\includegraphics[width=7.6cm]{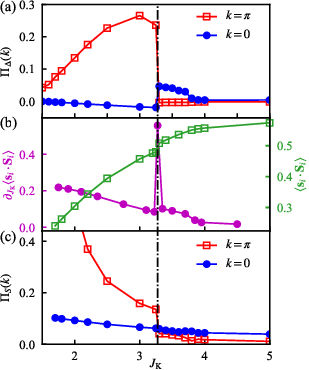}
	\end{center}
	\caption{Evolution of (a) the pair correlation $\Pi_\Delta(k)$ at $k=0$, $\pi$, (b) the local Kondo entanglement $\langle \mathbf{s}_i \cdot \mathbf{S}_i\rangle$ and its derivative with respect to $J_\mathrm{K}$, and (c) the spin correlation $\Pi_S(k)$ at $k=0$, $\pi$ as functions of the Kondo coupling $J_\mathrm{K}$ at $Q=0$. The vertical dash-dotted line represents the QPT at $J_\mathrm{K}=3.275$. Other parameters are $J_1=0.6$ and $n=0.8$.}
	\label{fig4}
\end{figure}

Interestingly, the quantum phase transition between the PDW and SC states is found to be first order. This is seen in Fig.~\ref{fig4}(a), where the PDW order parameter $\Pi_\Delta(k=\pi)$ experiences a sudden drop to zero at the transition point $J_\mathrm{K} = 3.275$, while simultaneously the SC order parameter $\Pi_\Delta(k=0)$ emerges and jumps to a finite value. The first-order transition is accompanied by a sudden jump of the Kondo correlation $\langle \mathbf{s}_i \cdot \mathbf{S}_i \rangle = \frac{\partial E_\mathrm{g} }{\partial J_\mathrm{K}}$ and the divergence of its derivative with respect to $J_\mathrm{K}$, as shown in Fig.~\ref{fig4}(b). This is distinctly different from the Ising universality observed in a related model studied via the bosonization \cite{Jaefari2012Fradkin, Mann2020Topology}.

The PDW-SC transition is closely associated with the change in the spin correlation function:
\begin{equation}{\label{eq_Sx}}
\Pi_S(x) = \left\langle \mathbf{S}_x\cdot \mathbf{S}_0 \right\rangle - \left\langle \mathbf{S}_x\right\rangle\cdot \left\langle \mathbf{S}_0 \right\rangle.
 \end{equation}  
The results in real space are given in Fig.~\ref{fig3}(c) and show exponential decay with distance for both $J_\mathrm{K} = 3.25$ and 3.3. In momentum space, as shown in Fig.~\ref{fig4}(c), the spin correlation function is dominated by the peak at $k=\pi$ for small $J_\mathrm{K}$ in the PDW region. At the transition $J_\mathrm{K}=3.275$, the peak suddenly drops to the same magnitude as $k=0$, so that the spin spectra shown in Fig. \ref{fig3}(d) become more flat over a wide momentum range in the uniform SC region. This suggests that the PDW order is closely related to the strong AFM spin fluctuations near the AFM phase boundary. Increasing the Kondo coupling reduces the spin fluctuations at $k=\pi$ and destroys the PDW order, but the fluctuations still remain sufficient to induce the uniform SC. Similar PDW-SC transition is found in EuRbFe$_4$As$_4$ \cite{Zhao2023Smectic}, where it is accompanied by the suppression of a helical magnetic order with increasing temperature. The triplet-pair correlations are also examined and excluded due to exponential decay with the distance. 

We therefore conclude that the intermediate quantum critical phase of the 1D Kondo-Heisenberg model may become unstable towards superconductivity owing to the AFM spin fluctuations, and a nonuniform PDW state may appear close to the AFM QCP. Very recently, the PDW state was discovered in the heavy fermion superconductor UTe$_2$ \cite{Aishwarya2023UTe2, Gu2023UTe2}, which is formed of two-leg ladders with frustrated magnetic interactions \cite{Xu2019PRL}.

\subsection{The Luther-Emery liquids}
We show here that both PDW and SC belong to a class of states known as the Luther-Emery liquids (LELs), which are characterized by the presence of a finite spin gap, the absence of a charge gap, and a power-law decay of the superconducting gap with system size \cite{Luther1974Backward, Orignac2003Superconducting}.

The power-law decay of the superconducting gap is obvious owing to the presence of PDW and SC and obeys $\Delta_\mathrm{SC}\propto L^{-K_\Delta}$. The finite spin gap can be seen from the spin correlation function plotted in Fig.~\ref{fig3}(c), where it decays exponentially as $|\Pi_S(x)| \propto \mathrm{e}^{-\Delta_S x}$ in both the PDW and SC states. The value of $\Delta_S$ reflects the magnitude of the spin gap.

We now demonstrate the absence of charge gap by calculating the charge density correlation function:
\begin{equation}{\label{eq_nx}}
\Pi_n(x) = \left\langle n_x n_0 \right\rangle - \left\langle n_x\right\rangle \left\langle n_0 \right\rangle,
\end{equation}
where $n_l=\sum_\sigma c^\dagger_{l\sigma} c_{l\sigma}$ is the electron density at site $l$. As shown in Fig.~\ref{fig3}(e), $\Pi_n(x)$ exhibits power-law decay with distance, $|\Pi_n(x)| \propto |x|^{-K_n}$, in both PDW and SC phases. In momentum space as shown in Fig. \ref{fig3}(f), both spectra are dominated by sharp peaks at $k_\mathrm{CDW}=0.2\pi~ (\mathrm{mod } ~2\pi)$. This implies a coexisting CDW order of the wave length $\lambda_\mathrm{CDW}=10$ in both phases. The much larger wave length indicates that this CDW order is not induced by the PDW state with $\lambda_\mathrm{PDW}=2$.

\begin{figure}[t]
	\begin{center}
		\includegraphics[width=8.2cm]{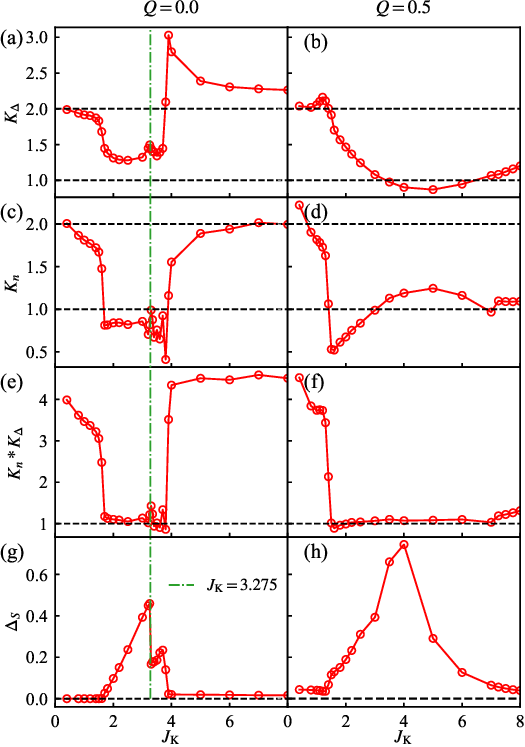}
	\end{center}
	\caption{Evolution of (a)(b) $K_\Delta$, (c)(d) $K_n$, (e)(f) $K_n*K_\Delta$, and (g)(h) the spin gap $\Delta_S$ as functions of the Kondo coupling $J_\mathrm{K}$ for $Q=0$ (left) and $0.5$ (right). The vertical dash-dotted line represents the PDW-SC transition point at $Q=0$. There is no PDW state at $Q=0.5$. Other parameters are $J_1=0.6$ and $n=0.8$.}
	\label{fig5}
\end{figure}  

The Luther-Emery liquids require a further constraint on the power-law exponents of the quasi-long-range singlet-pair correlations and charge density correlations, namely, $K_\Delta*K_n\simeq 1$. To see this, we plot in Fig.~\ref{fig5} the evolution of both parameters as well as the spin gap $\Delta_S$ as a function of the Kondo coupling $J_\mathrm{K}$ for $Q=0$ (left) and 0.5 (right). Both parameters decreases rapidly to a value smaller than 2 after the AFM or VBS order is suppressed at large $J_\mathrm{K}$, where $\Delta_S$ starts to take a non-zero value. Theoretically \cite{Giamarchi2003Quantum}, it is known that as the temperature $T\to0$, the pair susceptibility obeys $\chi_p\propto T^{-(2-K_\Delta)}$. $K_\Delta<2.0$ thus implies divergence of the pair correlation and a quasi-long-range order at zero temperature \cite{Jiang2022Pair, Zhang2022Pair}, which emerges together with the finite spin gap and the quasi-long-range density correlation. Around the QCP, $K_\Delta$ can also be slightly smaller than $2.0$ inside the AFM or VBS state, indicating the possible coexistence of a secondary PDW order. Obviously, increasing the frustration extends the range of the SC state and enhances the strength of the pair correlation as reflected in the smaller minimum of the $K_\Delta$ value in Fig.~\ref{fig5}(b). For $Q=0.5$, as depicted in the phase diagram Fig.~\ref{fig1}(b), the PDW state is weakened and disappears in the whole parameter range. Additionally, the relative strength of the charge density correlation and the pair correlation also depends on the magnitude of $J_1$ (not shown), revealing a subtle competition between CDW and SC, which has been intensively discussed in recent literatures \cite{Zhang2022Pair, Jiang2022Pair, Peng2021Gapless, Chen2021Double, Wandel2022Enhanced}. Nevertheless, over a wide range of the Kondo coupling, as shown in Figs.~\ref{fig5}(e) and \ref{fig5}(f), we find $K_\Delta*K_n \simeq 1$, in good agreement with the property of the Luther-Emery liquids \cite{Zhang2022Pair, Orignac2003Superconducting}.

\subsection{The Luttinger liquid}
For sufficiently large $J_\mathrm{K}$, the superconductivity is completely suppressed and the ground state becomes a Luttinger liquid. As is shown for $J_\mathrm{K}=5.0$ in Fig.~\ref{fig3}(a), $\Pi_\Delta (x)$ exhibits a rapid power-law decay with $K_\Delta>2$, indicating the absence of quasi-long-range pair correlation. The spectra $\Pi_\Delta(k)$ in momentum space are plotted in Fig.~\ref{fig3}(b) and are close to zero in the whole Brillouin zone. For $Q=0$, $K_\Delta$, $K_n$, and $K_\Delta*K_n$ all show a rapid change at $J_\mathrm{K}\approx 3.8$. Correspondingly, $\Delta_S$ vanishes as shown in Fig.~\ref{fig5}. These suggest a significant reduction in the charge density correlation and the pair correlation and confirm the breakdown of the Luther-Emery liquids in the region of large $J_\mathrm{K}$. In the meanwhile, the spin correlation function $\Pi_S(k)$ displays a plateau in momentum space and a small peak at $k=\pm 0.2\pi ~(\mathrm{mod } ~2\pi)$, resembling that of electrons with the Fermi wave vector $k^\mathrm{L}_\mathrm{F}=0.9\pi$. We thus conclude that the large $J_\mathrm{K}$ region is a Luttinger liquid (LL) with a large Fermi volume, consistent with previous calculations \cite{Khait2018Doped, Eidelstein2011Interplay, Shibata1996Friedel, Shibata1997Friedel, Shibata1999The}. Different from the first-order PDW-SC transition, the SC-LL transition is continuous. For $Q=0$, this is evidenced in Fig.~\ref{fig4} by the gradual disappearance of $\Pi_\Delta(k)$ with increasing $J_\mathrm{K}$ and the smooth evolution of the first and second derivatives of $E_\mathrm{g}$ at $J_\mathrm{K}\approx 3.8$.

\section{Discussion}{\label{SecDiss}}

\subsection{Evolution of the Fermi volume}{\label{SecFS}}  
 
 \begin{figure}[t]
 	\begin{center}
 		\includegraphics[width=8.6cm]{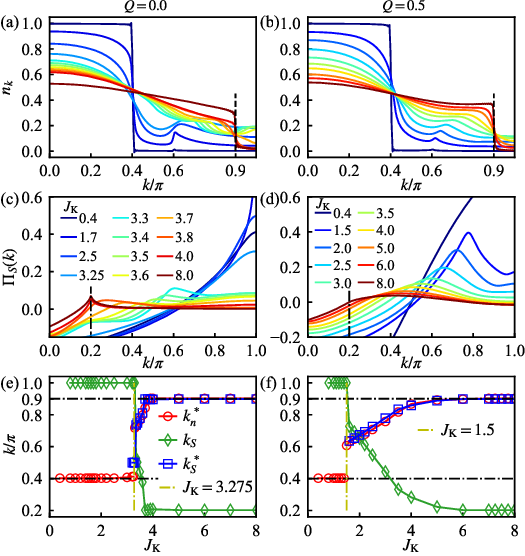}
 	\end{center}
 	\caption{Momentum distribution of (a)(b) the electron occupation $n_k$ and (c)(d) the spin correlation function $\Pi_S(k)$ for different values of $J_\mathrm{K}$ at $Q=0$ (left) and 0.5 (right). (e)(f) Evolution of the characteristic momentum $k^*_{n}$ (red circles), $k_{S}$ (green diamond), and $k^*_{S}$ (blue squares) as functions of $J_\mathrm{K}$ for $Q=0$  (left) and $0.5$ (right). The vertical dot-dashed lines in (e) and (f) represent the quantum phase transition into the SC. Other parameters are $J_1=0.6$ and $n=0.8$. }
 	\label{fig6}
 \end{figure} 

The PDW and SC states or the Luther-Emery liquids may be viewed to arise from pair instability of certain intermediate quantum critical phase between AFM or VBS of a small Fermi volume and the Luttinger liquid of a large Fermi volume. To understand the nature of this intermediate phase and resolve the theoretical controversial, we analyze below how the Fermi volume evolves in different regions of the phase diagram. This is done by defining a characteristic momentum $k^*_{n}$ where a jump or power-law singularity occurs in the momentum distribution of the electron density \cite{Moukouri1995Density, Tsunetsugu1997The, McCulloch2002Interplay, Xie2015Interplay, Eidelstein2011Interplay}:
\begin{equation}
n_k=\frac{1}{L} \sum_{l\sigma} \langle c^\dagger_{0\sigma} c_{l\sigma} \rangle \mathrm{e}^{-ikl}.
\end{equation}
Some typical results are given in Fig.~\ref{fig6}(a) for $Q=0$, 0.5 and $n=0.8$. The momentum distribution shows a sharp jump at $k^*_n=0.4\pi=k^\mathrm{S}_\mathrm{F}\equiv\frac{n}{2}\pi$ for $J_\mathrm{K}=0.4$ and $k^*_n=0.9\pi=k^\mathrm{L}_\mathrm{F}\equiv\frac{1+n}{2}\pi$ for $J_\mathrm{K}=8.0$, indicating the existence of a small Fermi volume in the weak Kondo coupling limit and a large Fermi volume in the strong coupling limit. At intermediate $J_\mathrm{K}$, a small peak appears around $k^*_n=0.6\pi=(1-\frac{n}{2})\pi$ due to the interaction with AFM spin correlations. In Fig.~\ref{fig6}(c), the spin spectra also exhibits an additional peak around similar momentum. Beyond $J_\mathrm{K}=3.275$, where the PDW order turns into the uniform SC state, a kink feature shows up in $n_k$ at a larger momentum, which evolves gradually with increasing Kondo coupling.

The overall evolution of $k^*_n$ with $J_\mathrm{K}$ is shown (red circiles) in Fig.~\ref{fig6}(e), where it jumps from $k^\mathrm{S}_\mathrm{F}=0.4\pi$ to around $k^*_{n}=0.7\pi$ at the PDW-SC transition $J_\mathrm{K}=3.275$ and then continuously increases to $k^\mathrm{L}_\mathrm{F}=0.9\pi$ at $J_\mathrm{K}=3.8$, indicating an intermediate region between two limits with well-defined small or large Fermi volumes. Interestingly, the PDW state is found to always have a small Fermi volume, reflecting a competition between the PDW order and the Kondo delocalization owing to the presence of strong AFM spin correlations. The sudden jump of the Fermi volume at the PDW-SC transition suggests that the PDW state might arise from a normal state with a medium Fermi volume. For $Q=0.5$, where the PDW order is gone, $k^*_{n}$ jumps at the VBS-SC transition and then increases continuously to $k^\mathrm{L}_\mathrm{F}$ as shown in Figs.~\ref{fig6}(b) and \ref{fig6}(f). The only difference is that for nonzero $Q$, the SC state persists even in the large Fermi volume region with very strong Kondo coupling as depicted in Fig.~\ref{fig1}(b). With increasing $Q$, the pair correlation strengthens and the region of the Luther-Emery liquids also expands. Consequently, the distance between the right boundary (the cyan solid line) of the SC state and the boundary (the yellow dotted line) at which the Fermi volume becomes large increases. In CeRhIn$_5$ \cite{Park2011}, superconductivity has also been observed to penetrate in the AFM and Fermi liquid regions.

\begin{figure}[t]
	\begin{center}
		\includegraphics[width=6.8cm]{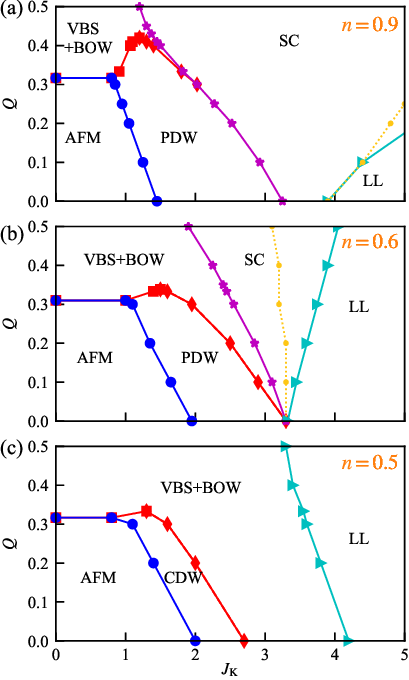}
	\end{center}
	\caption{Variation of the overall $Q$-$J_\mathrm{K}$ phase diagrams with the electron density: (a) $n=0.9$, (b) $n=0.6$, and (c) $n=0.5$ with $J_1=0.6$. The yellow dotted lines separate the regions with medium (left) and large (right) Fermi volume.}
	\label{fig7}
\end{figure} 

The above results are further supported by the analysis of the Friedel oscillation, which has been commonly used as an indicator of the size of the Fermi volume \cite{Shibata1996Friedel, Shibata1997Friedel, Shibata1999The, Khait2018Doped}. Unfortunately, our results show that the Friedel oscillation of the electron density is dominated by the wave vector $k_\mathrm{CDW}=0.2\pi$ of the CDW order and cannot be used to distinguish the small and large Fermi volume, so we have to consider the Friedel oscillation of the spins as reflected in the spin flucutation spectra $\Pi_S(k)$ for $Q=0$ in Fig.~\ref{fig6}(c) and $Q=0.5$ in Fig.~\ref{fig6}(d). In both cases, we find rich peak structures in the spin spectra. The peak positions $k_S$ as a function of the Kondo coupling are collected and compared in Fig. \ref{fig6}(e) for $Q=0$ and Fig. \ref{fig6}(f) for $Q=0.5$. In both cases, we see $k_S$ jumps from $\pi$ to an intermediate value at the PDW-SC or VBS-SC transition, and then gradually evolves to $0.2\pi = 2k^\mathrm{L}_\mathrm{F}~\mathrm{mod}~2\pi$ as $J_\mathrm{K}$ increases, resembling the behavior of $n_k$ in the whole SC region. The relation between $k_S$ and the Fermi wave vector agrees well with the expectation for a Luttinger liquid. By employing the same relation and defining $k^*_S$ via $k_{S} = \pm 2k^*_{S} ~(\mathrm{mod}~2\pi)$, we find that the derived $k^*_{S}$ is in quantitative agreement with $k^*_{n}$ even in the whole SC phase, which provides an additional support for the continuous evolution of the Fermi volume in the intermediate region. This is in contrast to earlier studies \cite{Eidelstein2011Interplay, Moukouri1995Density, Pivovarov2004}, which assume a continuous QPT between two phases with small or large Fermi volumes. The existence of an intermediate region without an apparent Fermi surface has also been observed previously in the literature  \cite{Eidelstein2011Interplay}, but they failed to notice the effect of CDW order in hiding the charge Friedel oscillation and concluded that $k^*_{n}$ and $k^*_{S}$ cannot represent a Fermi surface. Our results are consistent with recent proposal based on Schwinger boson calculations for frustrated Kondo lattices \cite{Wang2021Nonlocal, Wang2022Z2}.

\subsection{Variation of the pair instability}{\label{Secdoping}}  

  \begin{figure}[t]
	\begin{center}
		\includegraphics[width=7.2cm]{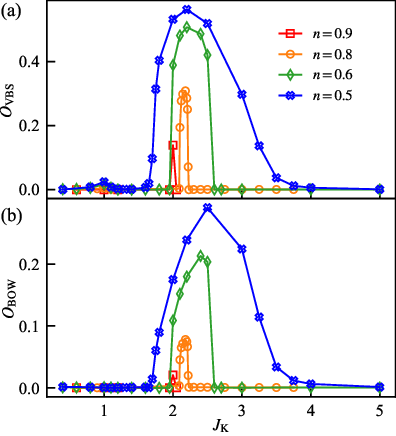}
	\end{center}
	\caption{Evolution of (a) $O_\mathrm{VBS}$ and (b) $O_\mathrm{BOW}$ at $Q=0.3$ for the electron density $n = 0.9$, $0.8$, $0.6$, and $0.5$ with $J_1=0.6$.}
	\label{AppBOWDoing}
\end{figure} 

The pair instability for PDW and SC may change with tuning parameters such as the electron density and the magnitude of $J_1$. Figure~\ref{fig7} shows the phase diagrams for $n=0.9$, 0.6, 0.5. Compared to Fig.~\ref{fig1}(b) for $n=0.8$, the AFM phase boundaries are only slightly changed since they are mostly determined by the Heisenberg interaction. The PDW and SC regions are suppressed with reducing $n$ and disappear at $n=0.5$, indicating substantial weakening of pair correlations. By contrast, the right VBS/BOW state is greatly enhanced at $n=0.5$ to cover a large portion of the phase diagram. This is also seen in their substantially enhanced order parameters in Fig.~\ref{AppBOWDoing}, which arise most probably due to the RKKY interaction mediated by itinerant electrons \cite{Huang2020Coupled, Xavier2003Dimerization, Xavier2008One}. Different from the antiferromagnetic $J_1$ and $J_2$ that may originate from the superexchange interaction, the induced RKKY interaction at $n=0.5$ is ferromganetic between nearest-neighbor spins and antiferromagnetic between next-nearest-neighbor spins. For sufficiently large $J_\mathrm{K}$, they may surpass $J_1$ and $J_2$ to give rise to the VBS/BOW order. Our calculations for smaller $J_1$ indeed confirm the suppression of pair correlations and the intermediate Luther-Emery liquids. For larger $n$, both the range and strength of the right VBS/BOW state reduce gradually, implying that its presence at other densities may be inherited from that at $n=0.5$. Ignoring the right VBS/BOW phase gives a much simpler phase diagram only with intermediate PDW and SC states.

\begin{figure}[t]
	\begin{center}
		\includegraphics[width=8.6cm]{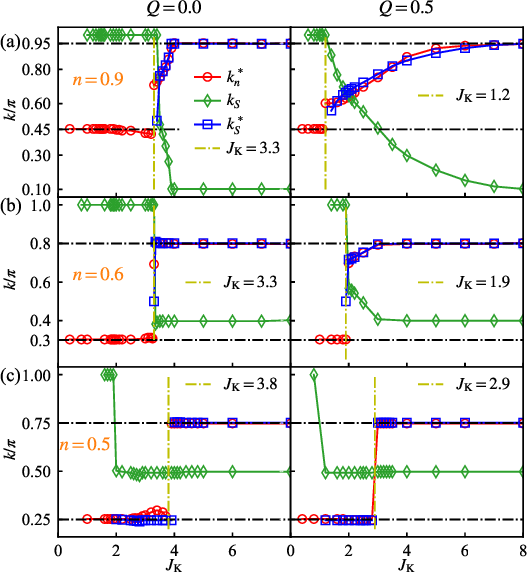}
	\end{center}
	\caption{Evolution of the characteristic momenta $k^*_{n}$ (red circles), $k_{S}$ (green squares), and $k^*_{S}$ (blue squares) for the electron density (a) $n=0.9$, (b) $n=0.6$, and (c) $n=0.5$ for $Q=0$ (left) and $0.5$ (right) with $J_1=0.6$. }
	\label{fig9}
\end{figure}  

As the PDW and SC regions diminish with decreasing $n$, the intermediate region with a medium Fermi volume also shrinks and eventually disappears as shown in Fig. \ref{fig9}. This leads to a sudden jump of the Fermi volume from small to large at the boundary of the VBS/BOW and Luttinger liquid states, which is detached from the AFM one. The lack of the medium Fermi volume region at small $n$ suggests a significant reduction in quantum fluctuations as electron density decreases. In fact, the spin gap $\Delta_S$ does decrease with decreasing $n$ or $Q$, indicating longer-range spin correlations due to the weakening of quantum fluctuations. In our 1D model, the intermediate phase exists even at $Q=0$ for large $n$. For higher dimension, frustration may be necessary in producing the quantum critical phase \cite{Wang2021Nonlocal, Wang2022Z2}.

\section{Conclusions}

To summarize, we performed systematic iDMRG simulations of the 1D Kondo-Heisenberg model with frustrated $J_1$-$J_2$ XXZ interactions among local spins and found rich ground states in the typical $Q$-$J_\mathrm{K}$ phase diagram. In particular, we found an intermediate region with strong pair correlations between the AFM or VBS/BOW of a small Fermi volume and the Luttinger liquid of a large Fermi volume. Apart from quarter filling, a long-range PDW order is observed close the the AFM phase boundaries, which turns into a uniform SC state for larger Kondo coupling through a first-order quantum phase transition with a sudden drop of AFM spin fluctuations. Analysis of the Fermi volume indicates a small one in the PDW state and a  continuous evolution in the SC state. It is suggested that the both may arise from a quantum critical normal state with medium Fermi volume. This implies a deep connection between the PDW, the unconventional superconductivity, and the non-Fermi liquid quantum critical phase. For $n$ close to 0.5 or weak frustration, both PDW and SC are suppressed and we find a sudden transition from small-to-large Fermi volumes. Our work provides a comprehensive understanding of the frustrated AFM Kondo lattice in one dimension. For higher dimension, similar mechanism may still play a role if quantum fluctuations are sufficiently strong with geometric frustrations or orbital degeneracy. It will be interesting to see if these may be verified in future studies using more sophisticated methods \cite{Chen2022Continuous}.

\begin{acknowledgments}
This work was supported by the National Key Research and Development Program of China (Grant No. 2022YFA1402203), the National Natural Science Foundation of China (NSFC Grant No. 12174429, No. 11974397), and the Strategic Priority Research Program of the Chinese Academy of Sciences (Grant No. XDB33010100).
\end{acknowledgments}

\appendix 
\section{Details on numerical simulations}
\begin{figure}[h]
	\begin{center}
		\includegraphics[width=8.6cm]{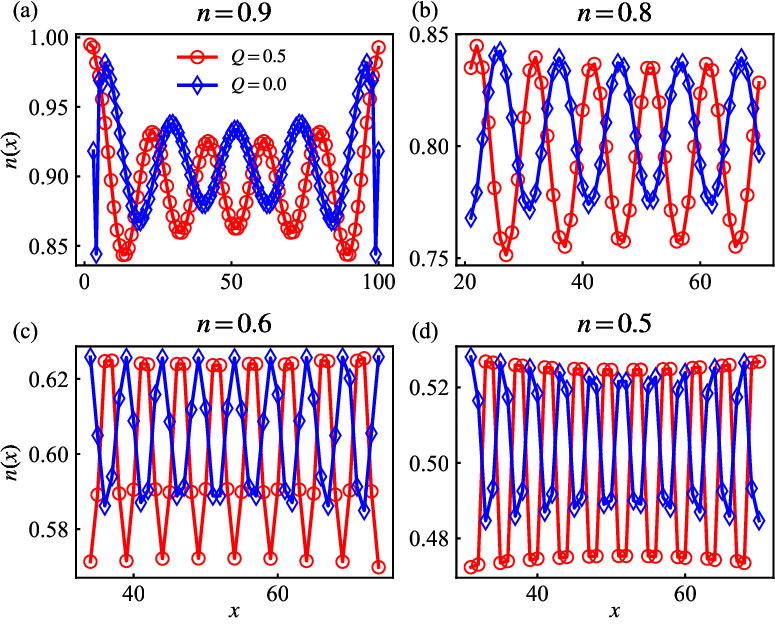}
	\end{center}
	\caption{Electron density $n(x)$ along with $x$ in the finite size DMRG simulation with $L=100$ for average electron densities $n=0.9$, 0.8, 0.6, 0.5. Other parameters are $J_\mathrm{K}=2.0$ and $J_1=0.6$. }
	\label{figS1}
\end{figure}

{\label{Appendix_A}}
We provide here more details on our numerical simulations. All data presented in the main text are obtained from iDMRG, and the finite size DMRG is only used to determine the unit cell size for iDMRG simulations. For example, as shown in Fig.~\ref{figS1} for the finite size system of length $L=100$ with the average electron density $n =0.9$, 0.8, 0.6, 0.5, the finite size DMRG calculations give the CDW wave length $\lambda_\mathrm{CDW}=20$, 10, 5, 4, respectively.  Taking into account the periods of PDW, AFM, VBS, and BOW, we choose the unit cell size to be $L_\mathrm{u}=20$,10,10, 8  in their respective iDMRG calculations. 

\begin{figure}[h]
	\begin{center}
		\includegraphics[width=8.6cm]{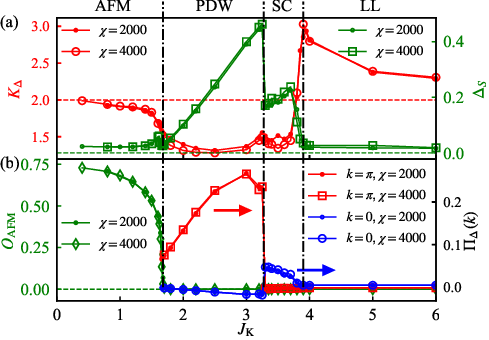}
	\end{center}
	\caption{The scanning process to obtain the ground states of the frustrated Kondo-Heisenberg model with $J_1=0.6$, $n=0.8$, and $Q=0.0$. }
	\label{figS2}
\end{figure}

\begin{figure}[h]
	\begin{center}
		\includegraphics[width=8.6cm]{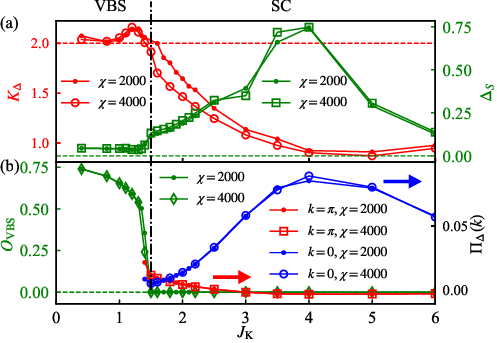}
	\end{center}
	\caption{The scanning process to obtain the ground states of the frustrated Kondo-Heisenberg model with $J_1=0.6$, $n=0.8$, and $Q=0.5$. }
	\label{figS3}
\end{figure}

To obtain the phase diagram for a given density $n$, we fix the frustration parameter $Q$, scan the Kondo coupling $J_\mathrm{K}$, calculate the quantities listed in Table~\ref{tab_1}, and fit the data to get the Luttinger parameter $K_\Delta$ and the spin gap $\Delta_{S}$. Taking $n=0.8$ as an example, two typical scanning processes are shown in Fig.~\ref{figS2} and Fig.~\ref{figS3} for $Q=0.0$ and 0.5, respectively. As shown in Fig.~\ref{figS2}, when $J_\mathrm{K}<1.6$, $O_\mathrm{AFM}$ has a finite value and $K_\Delta \ge 2$, so the ground state is AFM in this region. With increasing $J_\mathrm{K}$,  $K_\Delta$ reduces below 2.0 and the spin gap $\Delta_S$ obtains a finite value, indicating that the ground state becomes a  Luther-Emery liquid.  Depending on the position of the dominant peak in $\Pi_\Delta(k)$, this state is further divided into the PDW ($k=\pi$) and the SC ($k=0$). When $J_\mathrm{K}$ increases above 3.8, the spin gap $\Delta_{S}$ is gone and $K_\Delta\ge 2$, so the ground state is a Luttinger liquid. Similarly, for $Q=0.5$ shown in Fig.~\ref{figS3}, we find for $J_\mathrm{K}<1.5$ a VBS ground state with a finite $O_\mathrm{VBS}$ and $K_\Delta\ge 2$. While for $J_\mathrm{K}>1.5$, we find $K_\Delta<2.0$, a finite spin gap $\Delta_S$, and a dominant peak of $\Pi_\Delta (k)$ at $k=0$, implying a uniform SC ground state.

\begin{figure}[h]
	\begin{center}
		\includegraphics[width=8cm]{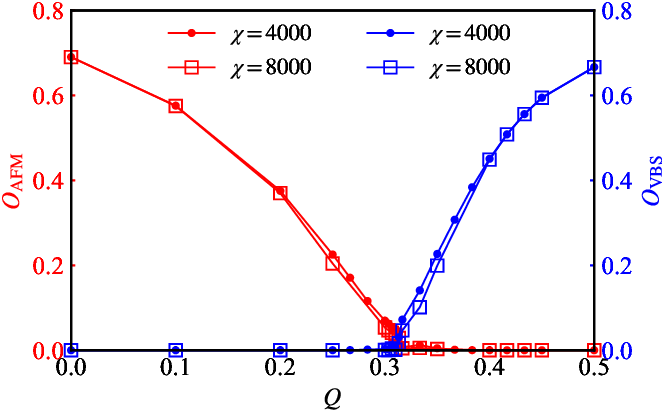}
	\end{center}
	\caption{Evolution of $O_\mathrm{AFM}$ and $O_\mathrm{VBS}$ with the frustration parameter $Q$ for virtual bond dimensions $\chi=4000$ and 8000. Other parameters are $J_\mathrm{K}=0.95$, $J_1=0.6$, and $n=0.8$.}
	\label{figS4}
\end{figure}

To determine the DQCP, we fix $J_\mathrm{K}$ and scan the frustration $Q$. As shown in Fig.~\ref{figS4} for $n=0.8$ and $J_\mathrm{K}=0.95$, $O_\mathrm{AFM}$ decreases continuously with increasing $Q$ and disappears at $Q_c=0.32$, at which $O_\mathrm{VBS}$ appears and increases continuously.

The results in the main text are all converged by comparing the results with different bond dimensions from $\chi=2000$ to $8000$. For example, in obtaining Fig.~\ref{fig1}(b) for $J_1=0.6$, $n=0.8$, and $Q=0.0$, we use  $U(1)_\mathrm{charge}\times U(1)_\mathrm{spin}$ symmetry and the virtual bond dimension $\chi=4000$. As show in Fig.~\ref{figS2}(b), the order parameters $O_\mathrm{AFM}$ for AFM, $\Pi_\Delta(k=0)$ for SC, and $\Pi_\Delta(k=\pi)$ for PDW with $\chi=4000$ converge well compared to that of $\chi=2000$. The calculations of the fitting parameters $K_\Delta$ and $\Delta_S$ are more challenging, but as shown in Fig.~\ref{figS2}(a), they also converge well. Similarly in Fig.~\ref{figS3} for $Q=0.5$, the results also converge with $\chi=4000$.  While for the DQCP shown in Fig.~\ref{figS4}, the results of $\chi=8000$ converge quite well with that of $\chi=4000$, which gives concrete evidence for the existence of the DQCP.

\end{document}